\documentclass{aa}
\usepackage{graphicx}

\begin{document}

\newcommand{\Msolar}{\mbox{${\; {\rm M_{\sun}}}$}}
\newcommand{\Zsolar}{\mbox{${\; {\rm Z_{\sun}}}$}}
\newcommand{\ha}{\hbox{H$\alpha$}}
\newcommand{\hb}{\hbox{H$\beta$}}
\newcommand{\hg}{\hbox{H$\gamma$}}
\newcommand{\hd}{\hbox{H$\delta$}}
\newcommand{\hi}{\hbox{H\,{\sc i}}}
\newcommand{\hii}{\hbox{H\,{\sc ii}}}
\newcommand{\oi}{\hbox{[O\,{\sc i}]}}
\newcommand{\oii}{\hbox{[O\,{\sc ii}]}}
\newcommand{\oiii}{\hbox{[O\,{\sc iii}]}}
\newcommand{\nii}{\hbox{[N\,{\sc ii}]}}
\newcommand{\sii}{\hbox{[S\,{\sc ii}]}}
\newcommand{\mgii}{\hbox{Mg\,{\sc i}+Mg\,{\sc h}}}
\newcommand{\nai}{\hbox{Na\,{\sc i}}}

\titlerunning{Spectroscopic study of BCGs}
\authorrunning{X.Kong \& F.Z.Cheng}

\title{Spectroscopic study of blue compact galaxies}
\subtitle{I. The spectra
\thanks{
Table 3 is presented in its entirety in the electronic form at 
the CDS via anonymous ftp to cdsarc.u-strasbg.fr (130.79.128.5) or via 
http://cdsweb.u-strasbg.fr/cgi-bin/qcat?J/A+A/.
Figure 4 is presented in its entirety in the on-line version of A\&A.
}
}

\author{X. Kong \inst{1,2,3}\thanks{E-mail: xkong@mpa-garching.mpg.de}
\and
F.Z. Cheng\inst{2,3}
}
\institute{
Max Planck Institute for Astrophysics, Karl-Schwarzschild-Str. 1, 
D-85741 Garching, Germany
\and 
Center for Astrophysics, University of Science and Technology 
of China, 230026, Hefei, P. R. China
\and 
National Astronomical Observatories, Chinese Academy of Sciences, 
100012, Beijing, P. R. China
}

\date{Received 19 February 2002; Accepted 2 May 2002}

\abstract{
Blue compact galaxies are compact objects that are dominated by 
intense star formation. Most of them have dramatically different 
properties compared to the Milky Way and many other nearby galaxies.
Using the IRAS, \hi\ data, and optical spectra, we wanted to measure 
the current star formation rates, stellar components, metallicities, 
and star formation histories and evolution of a large blue compact 
galaxy sample. We anticipate that our study will be useful as a 
benchmark for studies of emission line galaxies at high redshift. 
In the first paper of this series, we describe the selection, 
spectroscopic observation, data reduction and calibration, and 
spectrophotometric accuracy of a sample of 97 luminous blue compact 
galaxies.
We present a spectrophotometric atlas of rest-frame spectra, as well 
as tables of the recession velocities and the signal-to-noise ratios.  
The recession velocities of these galaxies are measured with an 
accuracy of $\delta$V $<$ 67 km s$^{-1}$. 
The average signal-to-noise ratio of sample spectra is $\sim$ 51. The 
spectral line strengths, equivalent widths and continuum fluxes are 
also measured for the same galaxies and will be analyzed in the next 
paper of this series. The atlas and tables of measurements will be 
made available electronically.

}

\maketitle

\section{Introduction}

Blue compact galaxies (BCGs) have luminosities in the range 
$ M_B \simeq -12 $ mag to $ M_B \simeq -21 $ mag (\cite{kunth00}).
Much work has been devoted to the study of BCGs, including image (
such as \cite{kunth88}; \cite{doublier99}; \cite{cairos01}), and 
spectra (such as \cite{kunth79}; \cite{terlevich91}; 
\cite{kinney93}; \cite{kunth97}) observation in all wavelengths.
These galaxies have an average surface brightness higher than 20 
mag arcsec$^{-2}$ (\cite{gordon81}), and are characterized by their 
compact morphology and blue rest-frame colors 
($ B-V <0.45, B-r < 0.6 $; \cite{pisano01}). 
Most of BCGs have a saturated region surrounded by no, or only small 
amounts of, nebulosity, and show no regular structures, such as spiral 
arms, but may display irregular features, such as jets, filaments, 
bridges, etc.(\cite{gordon81}). 
BCGs were first observed spectroscopically by Sargent \& Searle (1970). 
Their optical spectra show strong narrow emission lines superposed on an 
almost featureless continuum, similar to the spectrum of an \hii\ region. 
The least blue systems generally show both emission and absorption lines, 
while the bluest usually show only emission lines. The blue rest-frame 
colors and strong narrow emission line spectra indicate intense current 
star formation activity (\cite{cairos01}). 
Population synthesis models yield typical star formation rates between 1 
and 20 $\Msolar$ yr$^{-1}$ (\cite{mas99}; \cite{ostlin01}).

Neutral hydrogen observations, in the 21-cm line, of BCGs have shown 
that BCGs are typically rich in \hi\ gas, and there are indications 
that they have large \hi\ halos (\cite{thuan99}; \cite{pisano01}). 
Another interesting property found among BCGs is that some of them have 
significant under-abundance of elements heavier than helium.
The abundances of heavy elements in these galaxies range between 
$\Zsolar/50$ and $\Zsolar/2$, and put them among the least chemically 
evolved galaxies in the universe. The two most metal-deficient galaxies 
known, I Zw 18 ($\Zsolar/50$) and SBS 0335-052 ($\Zsolar/41$) are BCGs 
(\cite{doublier99}; \cite{izotov99}).
These  properties make BCGs represent an extreme environment for star 
formation, that differs from that in the Milky Way and in many other nearby 
galaxies. Detailed studies of these galaxies are not only important for 
understanding their intrinsic properties, but also crucial for understanding 
star formation processes, galaxy evolution and cosmological parameters 
(\cite{kunth00}; \cite{izotov01}).

The Hubble Space Telescope and the new generation of 8 m class telescopes 
have extended our knowledge of the early Universe by identifying galaxies 
down to magnitudes B$\sim$28 mag and redshifts z$\ge$3. Studies of 
intermediate-redshift galaxies have revealed a population of compact, 
luminous galaxies with high star formation rates. Jangren et al. (2002) have 
compared several of the brighter objects of some intermediate-redshift 
samples and found that most of the galaxies have small sizes, high 
luminosities (hence high surface-brightness), and very blue colors, similar 
to the BCGs (\cite{pisano01}). Therefore, the more accessible nearby BCGs 
should help us better understand these distant objects.

Despite the extensive work carried out during the past decade, age and star
formation histories of BCGs are not very constrained.
With the advent of modern detectors, many BCGs ($>$ 90\%) have been found to
contain an underlying old stellar population (Kunth \& \"{O}stlin 2000, 2001).
It suggests that most BCGs undergo a few or several short bursts of star 
formation followed by longer, more quiescent periods (\cite{kong99}). 
An important question is whether these BCGs are unique objects or not? 
Only a large spectral sample can help us answer this question. The data 
published so far are not sufficient to allow a systematic analysis 
of the star formation history of BCGs, because detailed and exhaustive 
studies have concentrated on particular type of BCGs, such as very 
metal-deficient galaxies.

Motivated by these facts, we have undertaken an extensive study of a 
large sample of BCGs, including the collction of IRAS, \hi\ and optical 
spectroscopy data. 
Our ultimate goal is to perform a detailed analysis of these galaxies, 
focusing on the aspects of the various stellar components, the star 
formation rates, the metallicities (chemical abundances), the age of the
underlying population, dust contents and star formation history.

In this paper, we present the blue compact galaxies sample, the 
optical spectroscopic observations and the method of reduction. 
The paper is organized as follows: in \S 2, we present the galaxy 
sample. In \S 3, we describe the spectral observations. The data 
reduction, and calibration are outlined in \S 4. The spectrophotometric 
atlas is presented in \S 5. In \S 6, we assess the data quality and 
spectrophotometric errors. Finally, some concluding remarks are given 
in \S 7. 
In the forthcoming papers of this series, we will study the constraints 
set by these observations on the physical properties of BCGs in our sample.

\section{Description of the sample}

BCGs are typically rich in \hi\ gas. Many large \hi\ surveys of BCGs 
were carried out, in particular by Gordon \& Gottesman (1981), 
Thuan \& Martin (1981), Thuan et al. (1999) and Smoker et al. (2000). 
Since we want to combine the optical spectra we obtain with HI data to 
constrain simultaneously the stars and gas contents of BCGs, we therefore
select our sample on the basis of these surveys.

Considering the observatory site, instruments and possible observation
times, we selected our sample according to the following criteria:
(1) $m_B^c < 17.0$ mag; 
(2) $\delta_{2000} > -12^{\circ}$; 
(3) $M_B^c < -17$ mag. 
Using these criteria, 83 BCGs were selected.
14 additional "dwarf"($M_B^c > -17$ mag) BCGs were also included because we
want to compare their properties with those of "luminous"($M_B^c < -17$ mag) 
BCGs.
The final sample consists of 97 BCGs. 
We emphasize that there is no guarantee that our sample has any statistical 
completeness.

Most (92/97) of the galaxies in our sample come from 
Gordon \& Gottesman (1981). These authors assembled a list of 99 blue 
compact galaxies from the Markarian, Haro and Zwicky lists. The \hi\ 
observations were made using the 91-m transit radio telescope of the 
National Radio Astronomy Observatory and the 305-m spherical radio 
telescope of the Arecibo Observatory. Three other galaxies in our sample 
come from Kinney et al. (1993). One galaxy comes from Thuan \& Martin (1981); 
these authors assembled a list of 115 blue compact dwarf galaxies 
(BCDs, $M_B > -18$ mag) known at that time from the objective prism 
surveys of Markarian and Haro, with a few objects from Zwicky and 
other investigators (we are presently observing most of these BCDs 
and will present our results in the future). Finally, one galaxy in our 
sample comes from Thuan et al. (1999). 

\setcounter{table}{0}
\begin{table*}[]
\caption{Global parameters of observed blue compact galaxies}
\begin{tabular}{lcclrrcl}
\hline
Galaxy&R.A.(2000)&Decl(2000)&m$_{\rm B}^c$&Dist.&M$_{\rm B}^c$
&E(B-V)$_G$&Other\\
Name&(h-m-s)&(d-m-s)&(mag)&(Mpc)&(mag)&&Name\\
\hline
Mrk335    & 00:06:19.5  &  +20:12:10  &13.72&102.0&-21.3 & 0.035 &PGC473  \\
IIIZw12   & 00:47:56.5  &  +22:22:23  &14.58& 80.0&-19.9 & 0.045 &Mrk347 \\
Haro15    & 00:48:35.4  &  -12:42:59  &13.48& 86.0&-21.2 & 0.023 &Mrk960  \\
Mrk352    & 00:59:53.3  &  +31:49:37  &14.24& 62.0&-19.7 & 0.061 &PGC3575 \\
Mrk1      & 01:16:07.2  &  +33:05:22  &14.69& 67.0&-19.4 & 0.060 &NGC449 \\
IIIZw33   & 01:43:56.5  &  +17:03:43  &14.56&109.0&-20.6 & 0.068 &Mrk360 \\
VZw155    & 01:57:49.4  &  +27:51:56  &15.00&111.0&-20.2 & 0.082 &Mrk364 \\
IIIZw42   & 02:11:33.5  &  +13:55:02  &14.51&107.0&-20.6 & 0.088 &Mrk366 \\
IIIZw43   & 02:13:45.0  &  +04:06:07  &14.05& 47.0&-19.3 & 0.042 &Mrk589 \\
VZw372    & 04:13:56.0  &  +29:09:28  &15.19R& 74.0&-19.2 & 0.709 &UGC2989 \\
IIZw18    & 04:38:39.7  &  +11:14:28  &16.41R& 59.0&-17.4 & 0.391 &PGC15715 \\
IIZw23    & 04:49:44.4  &  +03:20:03  &13.88&110.0&-21.3 & 0.063 &Mrk1087  \\
IIZw28    & 05:01:42.0  &  +03:34:28  &14.94&113.0&-20.3 & 0.060 &VV790B  \\
IIZw33    & 05:10:48.1  &  -02:40:54  &13.48& 36.0&-19.3 & 0.102 &Mrk1094 \\
IIZw35    & 05:16:59.3  &  +00:55:20  &16.90N& 95.0&-18.0 & 0.140 &PGC17037 \\
IIZw40    & 05:55:42.8  &  +03:23:30  &14.22&  9.0&-15.6 & 0.820 &UGCA116 \\
IIZw42    & 06:03:11.4  &  +07:49:37  &15.37& 69.0&-18.8 & 0.417 &UGC3393 \\
Mrk5      & 06:42:15.5  &  +75:37:33  &15.12& 13.3&-15.5 & 0.084 &UGCA130 \\
Mrk6      & 06:52:12.2  &  +74:25:37  &13.86& 76.0&-20.5 & 0.136 &UGC3547 \\
VIIZw153  & 07:28:12.0  &  +72:34:29  &13.40& 43.0&-19.8 & 0.030 &Mrk7 \\
VIIZw156  & 07:29:25.4  &  +72:07:44  &13.17& 50.0&-20.3 & 0.026 &Mrk8  \\
Haro1     & 07:36:56.4  &  +35:14:31  &12.39& 50.0&-21.1 & 0.043 &NGC2415 \\
Mrk385    & 08:03:28.0  &  +25:06:10  &14.57&110.0&-20.6 & 0.030 &PGC22615 \\
Mrk622    & 08:07:41.0  &  +39:00:15  &14.33& 94.0&-20.5 & 0.051 &UGC4229 \\
Mrk390    & 08:35:33.1  &  +30:32:03  &14.58&101.0&-20.4 & 0.041 &PGC24127 \\
Zw0855    & 08:58:27.4  &  +06:19:41  &14.50& 47.0&-18.9 & 0.069 &UGC4703 \\
Mrk105    & 09:20:26.3  &  +71:24:16  &15.88& 49.0&-17.6 & 0.066 &PGC26416 \\
IZw18     & 09:34:02.0  &  +55:14:28  &15.80& 10.9&-14.4 & 0.032 &Mrk116 \\
Mrk402    & 09:35:19.2  &  +30:24:31  &15.66R& 98.0&-19.3 & 0.020 &PGC27258 \\
IZw21     & 09:46:28.6  &  +45:45:09  &14.59& 67.0&-19.5 & 0.015 &UGC5225  \\
Haro22    & 09:50:11.0  &  +28:00:47  &14.94& 18.4&-16.4 & 0.025 &PGC28305 \\
Haro23    & 10:06:18.1  &  +28:56:40  &14.33& 17.5&-16.9 & 0.024 &UGCA201 \\
IIZw44    & 10:15:14.7  &  +21:06:34  &15.75& 81.0&-18.8 & 0.023 &PGC29934 \\
Haro2     & 10:32:31.9  &  +54:24:03  &13.15& 21.0&-18.5 & 0.012 &Mrk33 \\
Mrk148    & 10:35:34.8  &  +44:18:57  &14.65R& 96.0&-20.3 & 0.019 &UGC5747 \\
Haro3     & 10:45:22.4  &  +55:57:37  &12.81& 13.6&-17.9 & 0.007 &Mrk35 \\
Haro25    & 10:48:44.2  &  +26:03:12  &15.52&101.0&-19.5 & 0.033 &Mrk727 \\
Mrk1267   & 10:53:03.9  &  +04:37:54  &14.48& 77.8&-20.0 & 0.034 &Ark264$^{r1}$\\
Haro4     & 11:04:58.5  &  +29:08:22  &15.51&  7.9&-14.0 & 0.030 &Mrk36  \\
IZw26     & 11:25:36.2  &  +54:22:57  &16.80& 83.0&-17.8 & 0.014 &Mrk40  \\
Mrk169    & 11:26:44.4  &  +59:09:20  &13.95& 17.5&-17.3 & 0.014 &UGC6447 \\
Haro27    & 11:40:24.8  &  +28:22:26  &14.13& 24.0&-17.8 & 0.022 &Mrk1507 \\
Mrk198    & 12:09:14.1  &  +47:03:30  &14.56& 97.0&-20.4 & 0.017 &PGC38613 \\
IIZw57    & 12:09:32.9  &  +17:00:51  &13.60& 88.0&-21.1 & 0.036 &PGC38634 \\
Mrk201    & 12:14:09.7  &  +54:31:38  &12.86& 35.0&-19.9 & 0.015 &IZw33$^{r1}$\\
Haro28    & 12:15:46.1  &  +48:07:54  &13.21& 10.6&-16.9 & 0.016 &NGC4218 \\
Haro8     & 12:19:09.9  &  +03:51:21  &14.49& 18.6&-16.9 & 0.017 &Mrk49 \\
Mrk50     & 12:23:24.1  &  +02:40:45  &14.89R& 92.0&-19.9 & 0.016 &PGC40220 \\
Haro29    & 12:26:16.0  &  +48:29:37  &14.24&  4.8&-14.2 & 0.015 &IZw36 \\
\hline
\end{tabular}
\end{table*}

\setcounter{table}{0}
\begin{table*}
\caption{\it Continued}
\begin{tabular}{lcclrrcl}
\hline
Galaxy&R.A.(2000)&Decl(2000)&m$_{\rm B}^c$&Dist.&M$_{\rm B}^c$
&E(B-V)$_G$&Other\\
Name&(h-m-s)&(d-m-s)&(mag)&(Mpc)&(mag)&&Name\\
\hline
Mrk213    & 12:31:22.2  &  +57:57:52  &13.03R& 41.7&-20.1 & 0.012 &NGC4500$^{r1}$\\
Mrk215    & 12:32:34.7  &  +45:46:04  &14.47& 79.0&-20.0 & 0.017 &PGC41591 \\
Haro32    & 12:43:48.6  &  +54:54:02  &13.70& 67.0&-20.4 & 0.016 &IZw41 \\
Haro33    & 12:44:38.3  &  +28:28:19  &14.39& 12.5&-16.1 & 0.016 &UGCA294 \\
Haro34    & 12:45:06.6  &  +21:10:10  &14.86& 93.0&-20.0 & 0.041 &Ark386 \\
Haro36    & 12:46:56.4  &  +51:36:46  &14.67&  8.0&-14.8 & 0.015 &UGC7950 \\
Haro35    & 12:47:08.5  &  +27:47:35  &15.89R& 99.0&-19.1 & 0.013 &PGC43139  \\
Haro37    & 12:48:41.0  &  +34:28:39  &15.18& 57.0&-18.6 & 0.014 &Mrk444 \\
IIIZw68   & 12:58:02.4  &  +26:51:34  &14.34& 76.0&-20.1 & 0.009 &UGC8080 \\
IIZw67    & 12:58:35.2  &  +27:35:47  &13.69&101.0&-21.3 & 0.010 &NGC4853  \\
Mrk57     & 12:58:37.2  &  +27:10:34  &15.08&102.0&-20.0 & 0.014 &PGC44486$^{r2}$\\
Mrk235    & 13:00:02.1  &  +33:26:15  &15.07&100.0&-19.9 & 0.012 &PGC44694 \\
Mrk241    & 13:06:19.8  &  +32:58:25  &15.87&105.0&-19.2 & 0.014 &PGC45363  \\
IZw53     & 13:13:57.7  &  +35:18:55  &15.00R& 69.0&-19.2 & 0.011 &PGC45993 \\
IZw56     & 13:20:35.3  &  +34:08:22  &14.10& 93.0&-20.7 & 0.013 &UGC8387 \\
Haro38    & 13:35:35.6  &  +29:13:01  &14.87& 11.9&-15.5 & 0.012 &UGC8578 \\
Mrk270    & 13:41:05.7  &  +67:40:20  &13.61& 38.0&-19.3 & 0.020 &NGC5283  \\
Mrk275    & 13:48:40.5  &  +31:27:39  &14.52&106.0&-20.6 & 0.019 &PGC48992 \\
Haro39    & 13:58:23.8  &  +25:33:00  &14.49& 35.0&-18.2 & 0.016 &PGC49732 \\
Haro42    & 14:31:09.0  &  +27:14:14  &14.74& 60.0&-19.2 & 0.015 &Mrk685 \\
Haro43    & 14:36:08.8  &  +28:26:59  &15.08& 27.0&-17.1 & 0.020 &PGC52193 \\
Haro44    & 14:43:24.7  &  +28:18:04  &16.05R& 51.0&-17.5 & 0.017 &PGC52587 \\
IIZw70    & 14:50:56.5  &  +35:34:18  &14.04& 18.0&-17.2 & 0.012 &Mrk829 \\
IIZw71    & 14:51:14.4  &  +35:32:31  &13.34& 18.0&-17.9 & 0.013 &PGC53039 \\
IZw97     & 14:54:39.2  &  +42:01:26  &14.47& 36.0&-18.3 & 0.020 &PGC53299 \\
IZw98     & 14:55:15.6  &  +42:30:25  &13.91& 75.0&-20.5 & 0.019 &NGC5787 \\
IZw101    & 15:03:45.8  &  +42:41:59  &15.19& 68.0&-19.0 & 0.019 &Ark467 \\
IZw117    & 15:35:53.6  &  +38:40:37  &14.15& 77.0&-20.3 & 0.017 &UGC9922 \\
IZw123    & 15:37:04.2  &  +55:15:48  &15.44& 11.6&-14.9 & 0.014 &Mrk487 \\
VIIZw631  & 15:59:11.9  &  +20:45:31  &12.98& 62.0&-21.0 & 0.055 &NGC6027 \\
Mrk297    & 16:05:12.9  &  +20:32:32  &12.98& 65.0&-21.1 & 0.076 &NGC6052 \\
IZw147    & 16:23:12.0  &  +54:08:36  &15.80R& 75.0&-18.6 & 0.009 &PGC57975 \\
IZw159    & 16:35:21.1  &  +52:12:53  &15.70& 38.0&-17.2 & 0.029 &Mrk1499 \\
IZw166    & 16:48:24.1  &  +48:42:33  &15.03&106.0&-20.1 & 0.016 &Mrk499  \\
Mrk893    & 17:15:02.2  &  +60:12:59  &16.34& 83.7&-18.3 & 0.021 &1714+602$^{r3}$\\
IZw191    & 17:40:24.8  &  +47:43:59  &14.70R& 81.0&-19.8 & 0.019 &PGC60671 \\
IZw199    & 17:50:05.1  &  +56:40:27  &14.82& 74.0&-19.5 & 0.048 &PGC60950 \\
IZw207    & 18:31:10.4  &  +55:16:32  &15.31R& 79.0&-19.2 & 0.042 &PGC61982 \\
IIZw82    & 20:23:15.6  &  +00:39:52  &13.46& 56.0&-20.3 & 0.191 &UGC11546 \\
IVZw67    & 21:02:18.8  &  +36:41:44  &14.00N& 38.0&-18.9 & 0.280 &UGC11668 \\
IIZw172   & 22:14:45.9  &  +13:50:37  &13.23&108.0&-21.9 & 0.065 &NGC7237 \\
IVZw93    & 22:16:07.7  &  +22:56:33  &15.44R& 55.0&-18.3 & 0.069 &PGC68454 \\
Mrk303    & 22:16:26.8  &  +16:28:17  &14.62&104.0&-20.5 & 0.051 &NGC7244 \\
Zw2220    & 22:23:02.0  &  +30:55:29  &13.52& 93.0&-21.3 & 0.078 &UGC12011 \\
Mrk314    & 23:02:59.2  &  +16:36:19  &13.45& 31.0&-19.0 & 0.085 &Mrk314 \\
IVZw142   & 23:20:03.1  &  +26:12:58  &14.89&109.0&-20.3 & 0.086 &Mrk322 \\
IVZw149   & 23:27:41.2  &  +23:35:21  &12.66& 80.0&-21.9 & 0.043 &Mrk325 \\
Zw2335    & 23:37:39.6  &  +30:07:47  &15.37& 22.0&-16.3 & 0.093 &Mrk328 \\
\hline
\end{tabular}
\\r1: Kinney et al. (1993); r2: Thuan \& Martin (1981);
r3: Thuan et al. (1999);\\ 
The other BCGs were selected from Gordon \& Gottesman (1981)
\end{table*}

In total our sample contains 97 blue compact galaxies, the general 
parameters of which are listed in Table 1. Column (1) lists the 
galaxy name following Haro (\cite{haro56}), Markarian (Mrk, 
\cite{markarian89}), and Zwicky (Zw, {\cite{zwicky71}). Columns (2) 
and (3) list the right ascension and declination at epoch 2000, 
taken from NED \footnote{The NASA/IPAC Extragalactic Database (NED) 
is operated by the Jet Propulsion Laboratory, California Institute of 
Technology, under contract with the National Aeronautics and Space 
Administration.}. 
Column (4) lists the apparent blue magnitude ($m_B^c$) corrected for 
instrumental effects, extinction, inclination, and doppler velocity.  
Most magnitudes were taken from Gordon \& Gottesman (1981). The other 
galaxies were corrected for instrumental effects, extinction, inclination 
based on de Vaucouleurs et al. (1991, RC3) and NED, as indicated by "R",
"N".  
Column (5) lists the distance (D) to the galaxy, in megaparsecs, assuming 
a Hubble constant of $H_0$=75 km s$^{-1}$ Mpc$^{-1}$ and $q_0=0.5$. 
Column (6) lists the absolute blue magnitude ($M_B^c$,), derived from 
Column (4) and Column (5) with $M_B^c =m_B^c - 5{\rm log D} +5 $.  
Column (7) lists the Galactic foreground reddening, E(B-V).
The last column (8) lists other commonly used designations for the galaxies.

The spatial distribution of the 97 sample galaxies on the sky is 
show in Fig. 1. The absolute blue magnitude ($M_B^c$) distribution 
of the 97 sample galaxies is presented in Fig. 2. 

\begin{figure}
\centering
\includegraphics[angle=-90,width=\columnwidth]{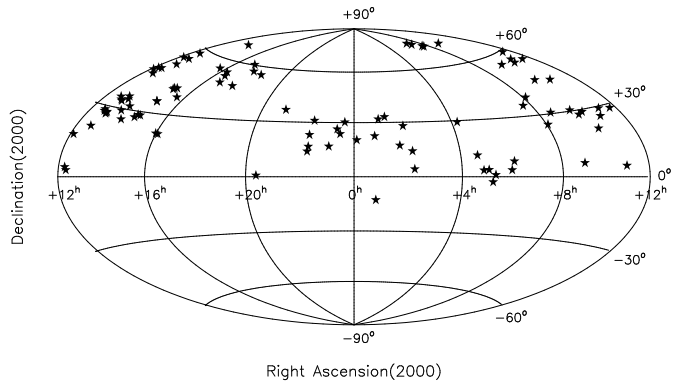}
\caption{The distribution of blue compact galaxies in the sky 
shown in equal-area projection, centered on Declination (latitude) 
0, Right Ascension (longitude) 0.}
\label{fig1}
\end{figure}

\begin{figure}
\centering 
\includegraphics[angle=-90,width=\columnwidth]{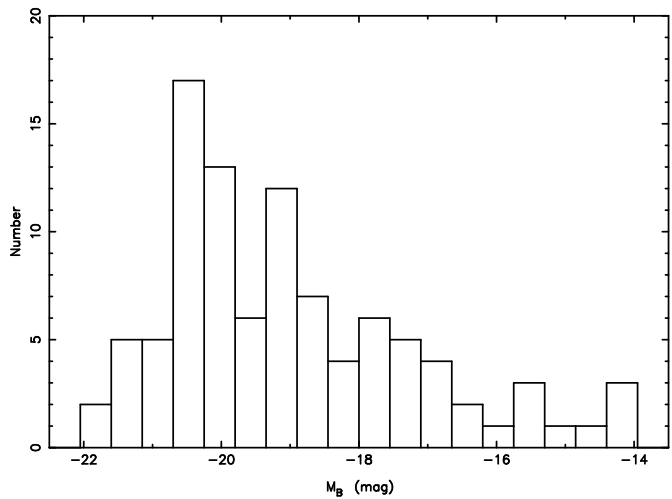}
\caption{The distribution of absolute blue magnitudes ($M_B$) of 
the 97 blue compact galaxies in our sample, 83 have $M_B < -17 $ 
mag.}
\label{fig2}
\end{figure}

\section{Spectroscopic observations}

Two-dimensional (long-slit) spectra were obtained using the OMR 
(Optomechanics Research Inc.) spectrograph mounted at the 
Cassegrain focus of the 2.16 m telescope at the XingLong Station of 
the Beijing Astronomical Observatory (BAO) in China, which also belongs 
to the National Astronomical Observatories, Chinese Academy of 
Sciences.  The data were acquired during 26 predominantly moonless 
nights between March 1997  and January 2002. The vast majority of 
the observations were obtained during transparent or photometric 
conditions, the remainder through thin clouds.

The OMR spectrograph was equipped with a TEKTRONIX TEK1024, AR coated 
back illuminated CCD with 1024 $\times $ 1024, 24 $\mu$m pixels.  
A 300 line mm$^{-1}$ grating blazed at 5500 {\AA} in first order was 
used, giving a dispersion of 4.8 \AA\,pixel$^{-1}$ and a total 
spectral coverage of $\sim$ 4500 \AA.  According to the condition 
of seeing, we adjusted the slit width in between 2$^{\prime\prime}$ and 
3$^{\prime\prime}$ on each night and used the selected configuration for 
all target spectra.  Our 
spectra were centered on 5500 \AA\ or 6000 \AA\ (see Table 2), at a 
resolution (FWHM) of $\sim$ 10 {\AA}, thus covering many interesting 
spectral features ranging from \oii$\lambda$3727 in the blue to 
\sii$\lambda$6731 in the red.

The slit was generally placed across the nucleus of each galaxy. 
In some cases where the location of the nucleus was not obvious, 
the slit was aligned to intersect the brightest part of the galaxy 
in order to maximize the chance of detecting the nucleus.  Otherwise, 
the spectrograph was rotated about the optical axis of the telescope 
as required to keep the slit approximately perpendicular to the 
horizon, to minimize loss of light due to differential atmospheric 
refraction (e.g., \cite{filippenko82}).

Dome flats, illuminated by a hot, spectroscopically featureless 
arc lamp, were taken at the beginning and end of each night. He-Ar or 
Fe-Ar (before 1998, see Column 6 of Table 2) lamps were observed 
immediately after observing each target object at the same position 
as the object for wavelength calibration. 

Two or more of the KPNO spectrophotometric standards from Massey 
et al. (1988), such as G191B2B, Hiltner~600, Feige~34, HZ~44, BD 
+28$^{\circ}$4211, were observed each night for absolute 
photometric calibration. In good conditions, a standard was 
observed about every 3 or 4 hours.
 
Table 2 lists the observations log and instrumental parameters for 
each run. Column (1) lists the observation run number, Column (2) 
the observation period, and Column (3) the number of nights actually
observed, i. e. that were not lost because of poor weather conditions. 
Column (4) lists the centering wavelength used for each run.  Columns 
(5) -- (8) list the slit width, the arc lamp used, the number of 
spectral obtained, and the standard stars for each run.  The individual 
integration times in each of the observations varied from 1200 
seconds to about an hour, depending on the magnitude of the object and
the seeing.  For 39 objects, observations were performed twice in 
order to get a higher signal-to-noise ratio spectrum, so that we 
have observed a total 136 BCGs spectra. 107 of the total 136 spectra were 
observed at airmass less than 1.2, only 4 spectra were observed at 
airmass larger than 1.5.  
Table 3 lists these observation parameters. 
Column (1) lists the object name, while Columns (2) -- (4) and (5) -- 
(7) list the observation date, exposure time, and airmass.

\begin{table*}
\caption{The observation date and instrumental parameters}
\begin{tabular}{clcccccl}
\hline
Obs.&Observation date&Number& cw.&slit&arc&Spec.& Standard stars\\
No.&&of nights&(\AA)& ($^{\prime\prime}$)&lamp&num.&\\
\hline
~1 &2002 Jan 08 -- Jan 12 & 5 & 5500 &  2.5 &    He/Ar   & 20& G191B2B Hilt600 Feige56 Feige25 HZ44\\
~2 &2001 Oct 12           & 1 & 5500 &  2.5 &    He/Ar   & ~4& Hilt600 HD192281 G191B2B BD174708\\
~3 &2001 Feb 24 -- Mar 01 & 4 & 5500 &  2.5 &    He/Ar   & 33& G191B2B Feige34 BD332642  HZ44 \\ 
~4 &2000 Apr 29 -- May 01 & 3 & 6000 &  2.3 &    He/Ar   & 24& Feige34 BD332642 HZ44 \\
~5 &2000 Feb 25           & 1 & 6000 &  2.2 &    He/Ar   & ~4& Hilt600 Feige34 \\
~6 &2000 Jan 01 -- Jan 02 & 2 & 5500 &  3.0 &    He/Ar   & 17& G191B2B Hilt600 Feige34\\ 
~7 &1999 Aug 03 -- Aug 09 & 2 & 5500 &  2.0 &    He/Ar   & ~4& BD284211 HD192281 \\
~8 &1998 Aug 14 -- Aug 20 & 4 & 5500 &  2.5 &    He/Ar   & 15& BD284211 Kopff27 Feige15 BD332642 \\  
~9 &1998 Feb 17           & 1 & 5500 &  2.0 &    He/Ar   & ~2& Hilt600 Feige98 \\
10 &1997 Mar 18 -- Mar 20 & 3 & 6000 &  2.0 &    Fe/Ar   & 13& Feige34 Feige98 Sa29130\\
\hline
\end{tabular}
\end{table*}

\section{Data reduction}

The spectroscopic reductions were made using the Image Reduction 
Analysis Facility (IRAF)\footnote[1]{IRAF is 
distributed by the National Optical Astronomy Observatories, which 
are operated by the Association of Universities for Research in 
Astronomy, Inc., under cooperative agreement with the National 
Science Foundation.} packages CCDRED, TWODSPEC and ONEDSPEC.  
For each night of data, the following steps 
were performed: 
(a) interpolation over bad columns, dead and hot pixels; 
(b) bias subtraction; 
(c) division of each frame by a flat-field exposure to remove 
multiplicative gain and illumination variations across the chip; 
(d) extraction of one dimensional spectra for each observation from 
the two-dimensional image by summing the pixels within the aperture 
at each point along the dispersion axis and subtracting out the sky 
background; 
(e) wavelength calibration and subsequent resampling of the data on 
a linear wavelength grid; 
(f) flux calibration of the extracted spectra, using flux standard 
stars from the KPNO standards sample; 
(g) correction for extinction using a standard atmospheric extinction 
law.

\subsection{Basic reductions}

Pixel-to-pixel variations in the response were removed through 
division by appropriately normalized exposures of the dome 
illuminated by a hot, spectroscopically featureless lamp.  The dark 
counts were so low that their subtraction was not performed.  Cosmic 
ray events in each input image were detected and replaced by the 
average of the four neighbors with the IRAF task {\it cosmicrays}. 
The remaining cosmic ray hits were flagged manually and were 
subsequently removed by interpolation.

We extracted the spectra using objectively defined apertures.  The 
peak of the galaxy light distribution was used to trace the 
extraction aperture. For the object spectra, we used a fixed 
aperture, 6 pixels along the slit centered on the brightest pixel 
in the portion of the spectrum between 5300--5600\AA.  The 
background sky level was determined from areas as close to the 
nucleus as possible, taking care not to include the contribution 
from extended emission near the nucleus.  
One-dimensional spectra of the standard stars were extracted in 
exactly the same manner as for the galaxy spectra, using an effective 
slit length to contain all the stellar light.

Wavelength calibration was carried out by fitting a cubic spline 
to unblended emission lines of He, Ar, and Fe in the comparison lamp 
spectra. These spectra were also used to measure the spectral 
resolution as a function of position on the CCDs. More than 20 lines 
were used to establish the wavelength scale.  Typical rms residuals 
in the cubic spline fits were 0.3 \AA. The accuracy of the wavelength 
calibration was better than 1.5 \AA.

The extracted spectra were flux calibrated on a relative flux scale 
using more than two KPNO standard stars. Cubic spline sensitivity 
functions of ninth order were fit interactively for each of the 
standard star observations. The sensitivity function relates the 
measured intensity to the (calibrated) flux density (in 
ergs\,s$^{-1}\,$cm$^{-2}\,$\AA$^{-1}$) as a function of wavelength, 
after removing atmospheric extinction.

Atmospheric extinction was corrected using the mean extinction 
coefficients for the Xinglong station (BAO), that were measured in 
the Beijing-Arizona-Taiwan-Connecticut (BATC) multi-color survey 
(\cite{kong00}).  There is little error introduced by this 
procedure, since the observations were restricted to small air 
masses, usually less than 1.2 and always less than 1.7.

The telluric O$_2$ absorption lines near 6280 and 6860\AA\ (the "B 
band") were removed through division by normalized, intrinsically 
featureless spectra of the standard stars. Large residuals caused 
by mismatches at the sharp, deep band were not removed.

\subsection{Additional reductions}

To measure the rest-frame spectral line properties of the galaxies, 
we first measured the recession velocity of each galaxy by 
averaging the recession velocities, $V_{o} =c \Delta 
\lambda / \lambda_0$ of different lines, where $\lambda_0$ is the 
rest-frame wavelength of the line. 

For galaxies with emission lines, recession velocities
were determined from the average of five measurable emission lines, 
\oii$\lambda$3727, \hb$\lambda$4861, \oiii$\lambda$4959, 
\oiii$\lambda$5007, \ha$\lambda$6563. 
For those objects without emission lines, velocities were obtained 
from the average of 5 measurable absorption lines, Ca\,K$\lambda$3935, 
\hd$\lambda$4101, G\,band$\lambda$4306, \mgii$\lambda$5177 and 
\nai$\lambda$5896.
For the galaxy, IVZw\,67, that has neither strong emission nor 
absorption lines, we adopted the redshift listed by NED.
The accuracy of the recession velocities in our sample ranges 
from very good (3 -- 55 km s$^{-1}$ uncertainties) when the 
galaxy has strong emission lines, to relatively poor (7 -- 67 km 
s$^{-1}$ uncertainties) when the galaxy has only absorption lines. 
The recession velocity distribution for all the sample galaxies with 
spectroscopic observations from this study are presented in Figure 3. 
All galaxies in the sample have velocities below 9000 km s$^{-1}$. On 
the basis of these recession velocities , the calibrated spectra were 
shifted to rest-frame wavelengths.
Column (8) -- (9) of Table 3 list the galaxy recession velocities in 
km s$^{-1}$, the uncertainty of the recession velocity.

\begin{figure}
\includegraphics[angle=-90,width=\columnwidth]{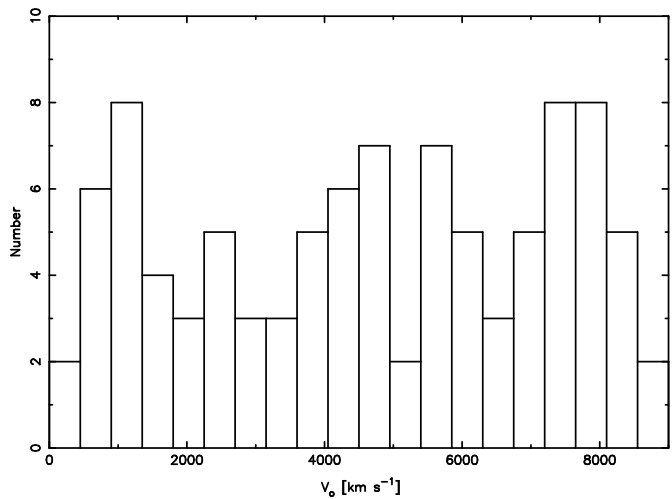}
\caption{Historgram of the recession velocity distribution
for all the 97 sample galaxies}
\label{fig3}
\end{figure}

The foreground reddening cause by our Galaxy was corrected using the 
values in Column (7) of Table 1. The wavelength dependence of the 
extinction was assumed to follow the empirical selective extinction 
function of Cardelli et al. (1989), with $R_V=A_V/E(B-V)=3.1$.

The rest frame spectra were normalized to the average flux in a 
50\AA\ interval centered at 5500\AA\ and subsequently resampled on 
a uniform wavelength grid spanning the range 3580--7400\AA.

For each galaxy, multiple exposures taken with the same setting, 
sometimes over several epochs, were combined in a weighted average, 
with the weights determined by the signal-to-noise ratios.

\section{The spectral atlas}

The calibrated spectra of the nuclei of all survey 
galaxies are presented in Figure 4. The objects are ordered with 
increasing right ascension at the epoch 2000 ($\alpha _{2000}$), 
as in Table 1. The spectra have all been normalized to the flux in 
a 50\AA\ wavelength interval centered on 5500\AA.  The ordinate 
displays normalized flux, $F(\lambda)/F(5500)$, and the abscissa 
wavelength(\AA) in the restframe of the galaxy, binned linearly 
to 2 \AA. All spectra were plotted on a common wavelength scale of 
3580 -- 7400\AA.

The spectra are generally of high quality.  The signal-to-noise 
ratios (SNRs) of the spectra, measured at 5500\AA\ over a region 
between 5475 and 5525\AA, are high. 
The SNRs of 89 sample galaxies are larger than 20, the average 
over the sample galaxies being $\sim$ 51. 
Column (10) -- (11) of Table 3 list the flux and signal-to-noise 
ratios in a 50\AA\ wavelength interval centered on 5500\AA.

Because the original BCGs sample of Haro (1956), Markarian et al.(1989), 
and Zwicky (1971) were identified on the basis of their high surface 
brightness on photographic plates and strong UV emission, they do not all 
have spectra similar to those of \hii\ regions. We find that some BCGs 
have broad H$\alpha$ spectral lines, such as Mrk 6, Mrk 335, and other 
objects. In fact, based on V\'{e}ron-Cetty \& V\'{e}ron (2001), 
10 sample galaxies are found to be Seyfert galaxies. 13 of 97 BCGs 
have no emission lines.  The other 74 spectra are typical \hii\ 
region spectra, with strong narrow nebular emission lines.

\begin{table*}
\caption{Summary observations and measurements for our blue compact galaxies sample}
\begin{tabular}{lccccccrcrr}
\hline
Galaxy&
\multicolumn{3}{c}{\hrulefill\ 1$st$ observation \hrulefill}&
\multicolumn{3}{c}{\hrulefill\ 2$nd$ observation \hrulefill}&
V$_{\rm o}$&$\delta$V&
F$_{5500}$&SNRs\\
Name&Date&Exp.(s)&Air&Date&Exp.(s)&Air&(km s$^{-1}$)&(km s$^{-1}$)&&\\
\hline
Mrk335   & 14/08/1998& 4000.& 1.07& 02/01/2000& 1500.& 1.12&  7978.3& 21.4&  9.0192& 128.0 \\
IIIZw12  & 12/10/2001& 1800.& 1.06& 10/01/2002& 1800.& 1.13&  5670.2& 22.6&  2.4061&  87.3 \\
Haro15   & 19/08/1998& 2000.& 1.71& 09/08/1999& 1800.& 1.68&  6478.3& 42.1&  3.0558& 155.2 \\
Mrk 352  & 10/01/2002& 1800.& 1.12&           &      &     &  4314.8& 33.5&  2.8522&  30.0 \\
Mrk 1    & 09/01/2002& 1800.& 1.06&           &      &     &  4723.4& 32.0&  1.5423&  28.6 \\
IIIZw33  & 02/01/2000& 1800.& 1.10&           &      &     &  8375.7& 27.3&  8.1323&  52.4 \\
VZw155   & 02/01/2000& 1800.& 1.05&           &      &     &  8554.9& 42.4&  6.8131&  70.3 \\
IIIZw42  & 24/02/2001& 1800.& 1.55& 10/01/2002& 1800.& 1.22&  8262.3& 25.8&  2.9035&  64.8 \\
IIIZw43  & 25/02/2001& 1800.& 1.68&           &      &     &  4202.2& 35.2&  3.9608&  63.3 \\
VZw372   & 02/01/2000& 1800.& 1.02& 24/02/2001& 1800.& 1.11&  4249.3& 31.4&  19.849&  59.9 \\
IIZw18   & 25/02/2001& 1800.& 1.21&           &      &     &  4020.3& 66.5&  5.6460&  23.1 \\
IIZw23   & 24/02/2001& 1800.& 1.41& 27/02/2001& 1800.& 1.33&  8310.0& 36.9&  2.5924&  70.3 \\
IIZw28   & 25/02/2000& 2000.& 1.26& 25/02/2000& 2000.& 1.29&  8552.4& 17.8&  0.6340&  41.6 \\
IIZw33   & 02/01/2000& 2400.& 1.38&           &      &     &  2068.7& 30.3&  4.7074&  34.4 \\
IIZw35   & 25/02/2001& 1800.& 1.38& 08/01/2002& 1800.& 1.30&  7402.0& 46.8&  2.5327&  53.0 \\
IIZw40   & 25/02/2000& 1800.& 1.28& 25/02/2000& 1800.& 1.32&   770.5& 15.6&  5.8835&  47.2 \\
IIZw42   & 24/02/2001& 1800.& 1.24& 01/03/2001& 1800.& 1.20&  5146.6& 31.4&  6.0929&  34.5 \\
Mrk5     & 27/02/2001& 1800.& 1.22& 11/01/2002& 2400.& 1.23&   772.1& 34.4&  0.6563&  18.4 \\
Mrk6     & 01/01/2000& 1800.& 1.21&           &      &     &  5677.7& 24.9&  7.3775&  58.0 \\
VIIZw153 & 02/01/2000& 1800.& 1.18& 02/01/2000&  600.& 1.19&  2666.1& 19.6&  2.9714&  21.2 \\
VIIZw156 & 24/02/2001& 1800.& 1.18& 01/03/2001& 1800.& 1.20&  3630.4& 23.2&  0.9874&  19.5 \\
Haro1    & 20/03/1997& 2400.& 1.25& 17/02/1998& 2000.& 1.02&  4081.9& 26.0&  2.5661&  26.9 \\
Mrk385   & 01/01/2000& 1800.& 1.04&           &      &     &  8181.2& 13.3&  3.1053&  45.7 \\
Mrk622   & 02/01/2000& 1800.& 1.01&           &      &     &  6666.6& 28.3&  7.3464&  52.5 \\
Mrk390   & 02/01/2000& 1800.& 1.03&           &      &     &  7496.4& 23.7&  4.3751&  29.6 \\
Zw0855   & 27/02/2001& 1800.& 1.28&           &      &     &  3110.0& 14.7&  0.9041&  19.6 \\
Mrk105   & 01/05/2000& 1800.& 1.22& 09/01/2002& 1800.& 1.17&  3912.4& 42.7&  1.3456&  69.1 \\
IZw18    & 01/01/2000& 1800.& 1.03& 01/01/2000& 1800.& 1.04&   744.5& 10.1&  0.6808&  34.1 \\
Mrk402   & 02/01/2000& 1800.& 1.02&           &      &     &  7243.0& 12.6&  2.3288&  20.9 \\
IZw21    & 29/04/2000& 1800.& 1.10&           &      &     &  4874.0& 30.9&  2.6133&  46.1 \\
Haro22   & 24/02/2001& 1500.& 1.04& 24/02/2001&  900.& 1.03&  1131.0& 13.5&  1.0172&  25.8 \\
Haro23   & 24/02/2001& 1800.& 1.02&           &      &     &  1379.7& 10.2&  1.2867&  30.6 \\
IIZw44   & 24/02/2001& 1800.& 1.06& 08/01/2002& 1800.& 1.06&  6278.8& 34.1&  1.1679&  58.8 \\
Haro2    & 19/03/1997& 1200.& 1.03&           &      &     &  1583.9& 44.5&  4.9081&  65.8 \\
Mrk148   & 30/04/2000& 1800.& 1.00& 11/01/2002& 1800.& 1.01&  7318.0& 18.5&  1.1232&  34.3 \\
Haro3    & 18/03/1997& 1200.& 1.15& 01/01/2000& 1800.& 1.04&   904.7& 11.5&  6.6393& 110.3 \\
Haro25   & 30/04/2000& 1800.& 1.05&           &      &     &  7914.8& 13.4&  2.4676&  51.9 \\
Mrk1267  & 20/03/1997& 2400.& 1.30& 10/01/2002& 1800.& 1.24&  6185.1& 23.1&  2.8781&  53.4 \\
Haro4    & 01/05/2000& 1800.& 1.02&           &      &     &   740.4& 13.7&  0.7125&  19.0 \\
IZw26    & 01/05/2000& 2400.& 1.04&           &      &     &  6169.0& 20.7&  0.9970&  38.9 \\
Mrk169   & 17/02/1998& 3600.& 1.07& 29/04/2000& 1800.& 1.08&  1271.3& 55.4&  2.5631&  83.9 \\
Haro27   & 01/01/2000& 1800.& 1.02&           &      &     &  1800.1& 20.4&  4.4561&  50.1 \\
Mrk198   & 29/04/2000& 1800.& 1.06&           &      &     &  7358.9& 50.0&  2.6700&  43.1 \\
IIZw57   & 30/04/2000& 1800.& 1.09&           &      &     &  6627.2& 24.0&  4.9057&  70.2 \\
Mrk201   & 20/03/1997& 1800.& 1.04&           &      &     &  2619.0& 26.2&  8.0253&  55.2 \\
Haro28   & 01/05/2000& 1800.& 1.02&           &      &     &   743.3& 17.5&  1.0679&  25.6 \\
Haro8    & 30/04/2000& 2400.& 1.25&           &      &     &  1631.1& 13.7&  0.7415&  26.5 \\
Mrk50    & 24/02/2001& 1800.& 1.33&           &      &     &  7026.6& 24.6&  1.5518&  25.9 \\
Haro29   & 19/03/1997& 1800.& 1.01&           &      &     &   315.5& 17.1&  0.4020&  18.5 \\
\hline
\end{tabular}
\label{tab3}
\end{table*}

\setcounter{table}{2}
\begin{table*}
\caption{Continued}
\begin{tabular}{lccccccrcrr}
\hline
Galaxy&
\multicolumn{3}{c}{\hrulefill\ 1$st$ observation \hrulefill}&
\multicolumn{3}{c}{\hrulefill\ 2$nd$ observation \hrulefill}&
V$_{\rm o}$&$\delta$V&
F$_{5500}$&SNRs\\
Name&Date&Exp.(s)&Air&Date&Exp.(s)&Air&(km s$^{-1}$)&(km s$^{-1}$)&&\\
\hline
Mrk213   & 20/03/1997& 1800.& 1.06& 08/01/2002& 1800.& 1.10&  3377.5& 11.8&  4.8087&  57.7 \\
Mrk215   & 10/01/2002& 1800.& 1.03& 11/01/2002& 1800.& 1.02&  5986.9& 30.0&  6.1985&  80.8 \\
Haro32   & 25/02/2001& 1800.& 1.08&           &      &     &  4837.4& 11.3&  2.7195&  45.6 \\
Haro33   & 27/02/2001& 2400.& 1.05&           &      &     &  1000.1& 24.0&  0.8098&  25.8 \\
Haro34   & 30/04/2000& 1800.& 1.08&           &      &     &  7012.3& ~9.8&  2.0477&  54.5 \\
Haro36   & 01/05/2000& 2400.& 1.09&           &      &     &   379.8& 25.1&  0.3791&  18.2 \\
Haro35   & 30/04/2000& 1500.& 1.09&           &      &     &  7663.0& 15.8&  1.6945&  27.7 \\
Haro37   & 24/02/2001& 1800.& 1.02&           &      &     &  4213.4& ~2.8&  1.9236&  59.8 \\
IIIZw68  & 25/02/2001& 1800.& 1.07& 27/02/2001& 1800.& 1.04&  7241.2& 35.1&  2.3599&  55.3 \\
IIZw67   & 19/03/1997& 1500.& 1.03& 01/01/2000& 1500.& 1.04&  7624.1& 15.1&  6.0330& 109.1 \\
Mrk57    & 25/02/2001& 1800.& 1.03&           &      &     &  7718.2& 48.4&  0.7684&  27.2 \\
Mrk235   & 24/02/2001& 1800.& 1.01&           &      &     &  7079.7& 13.0&  1.3856&  38.4 \\
Mrk241   & 30/04/2000& 1800.& 1.08&           &      &     &  8053.9& 42.5&  1.3220&  48.1 \\
IZw53    & 09/01/2002& 1800.& 1.07& 09/01/2002& 1800.& 1.03&  4829.3& 38.9&  0.8252&  32.9 \\
IZw56    & 25/02/2001& 1800.& 1.01& 11/01/2002& 1800.& 1.04&  6880.9& 55.0&  1.1944&  65.1 \\
Haro38   & 27/02/2001& 1800.& 1.02&           &      &     &   981.3& 16.3&  1.2028&  33.3 \\
Mrk270   & 30/04/2000& 1500.& 1.17&           &      &     &  3112.3& 17.3&  5.3308&  93.3 \\
Mrk275   & 24/02/2001& 1800.& 1.01&           &      &     &  7653.4& 20.2&  1.3908&  56.9 \\
Haro39   & 27/02/2001& 1800.& 1.04&           &      &     &  2691.9& 14.4&  0.7904&  26.0 \\
Haro42   & 29/04/2000& 1800.& 1.10&           &      &     &  4605.6& 38.8&  1.3191&  73.2 \\
Haro43   & 27/02/2001& 1800.& 1.03&           &      &     &  1608.8& 10.1&  0.4845&  16.4 \\
Haro44   & 10/01/2002& 1800.& 1.21& 10/01/2002& 1800.& 1.14&  3600.8& 17.1&  0.6320&  28.6 \\
IIZw70   & 11/01/2002& 1800.& 1.11&           &      &     &  1126.4& 27.3&  1.8516&  47.7 \\
IIZw71   & 01/05/2000& 1800.& 1.01&           &      &     &  1225.2& 22.1&  0.7794&  23.1 \\
IZw97    & 24/02/2001& 1800.& 1.00& 27/02/2001& 1200.& 1.02&  2320.1& 52.3&  0.7185&  25.1 \\
IZw98    & 29/04/2000& 1800.& 1.07& 01/05/2000& 1200.& 1.03&  5478.8& 24.1&  4.2581&  53.7 \\
IZw101   & 24/02/2001& 1800.& 1.00&           &      &     &  4785.6& 30.0&  1.1713&  21.2 \\
IZw117   & 30/04/2000& 2400.& 1.01&           &      &     &  5602.5& 10.7&  2.2100&  52.9 \\
IZw123   & 18/03/1997& 1800.& 1.04& 20/03/1997& 2820.& 1.04&   584.0& 32.3&  0.9901& 132.2 \\
VIIZw631 & 01/05/2000& 1200.& 1.10&           &      &     &  4310.9& 22.6&  2.1508&  40.5 \\
Mrk297   & 19/03/1997& 1800.& 1.06&           &      &     &  4760.3& 37.6&  2.4518&  59.2 \\
IZw147   & 01/05/2000& 1800.& 1.06&           &      &     &  5370.3& 22.4&  1.3458&  29.8 \\
IZw159   & 20/08/1998& 1800.& 1.10& 20/08/1998& 1800.& 1.14&  2884.9& 13.4&  1.5788&  63.7 \\
IZw166   & 19/03/1997& 3000.& 1.02& 19/03/1997& 1200.& 1.01&  7721.4& 52.4&  3.3032& 148.0 \\
Mrk893   & 19/08/1998& 3600.& 1.07& 19/08/1998& 1800.& 1.12&  6267.9& 22.6&  0.5122&  50.3 \\
IZw191   & 29/04/2000& 1800.& 1.01&           &      &     &  5624.6& 27.8&  2.5024&  42.4 \\
IZw199   & 20/08/1998& 3000.& 1.10&           &      &     &  5485.6& 17.0&  2.2458&  40.8 \\
IZw207   & 12/10/2001& 1800.& 1.10&           &      &     &  5724.3& 22.1&  0.2198&  11.9 \\
IIZw82   & 15/08/1998& 1500.& 1.48&           &      &     &  3651.3& 18.3&  8.1681&  70.8 \\
IVZw67   & 15/08/1998& 1800.& 1.09& 03/08/1999&  600.& 1.02&  2550.0& -- &  272.96&  97.1 \\
IIZw172  & 12/10/2001& 1500.& 1.13&           &      &     &  7880.5& ~6.9&  4.0476&  38.9 \\
IVZw93   & 20/08/1998& 3600.& 1.11&           &      &     &  3595.7& 26.4&  1.1569&  19.2 \\
Mrk303   & 20/08/1998& 1800.& 1.10& 20/08/1998& 1800.& 1.09&  7437.3& 50.7&  1.6985&  88.8 \\
Zw2220   & 12/10/2001& 1800.& 1.03&           &      &     &  6829.8& 17.0&  2.9004&  58.8 \\
Mrk314   & 09/08/1999& 1500.& 1.09&           &      &     &  2079.8& 33.6&  5.5239&  61.0 \\
IVZw142  & 09/01/2002& 1800.& 1.25&           &      &     &  8109.9& 26.8&  0.8097&  22.8 \\
IVZw149  & 14/08/1998& 3000.& 1.06& 20/08/1998& 2500.& 1.05&  3224.4& 30.3&  2.4436&  37.8 \\
Zw2335   & 15/08/1998& 4000.& 1.05& 09/08/1999& 1800.& 1.02&  1338.1& 52.5&  3.1063& 113.5 \\
\hline
\end{tabular}
\\The absolute flux (10{\it th} column) is in units of $10^{-15} 
{\rm ergs s^{-1} cm^{-2}  \AA^{-1}}$.
\end{table*}
The atlas and tables of measurements will be made available 
electronically.

\setcounter{figure}{3}
\begin{figure*}
\includegraphics[width=\textwidth]{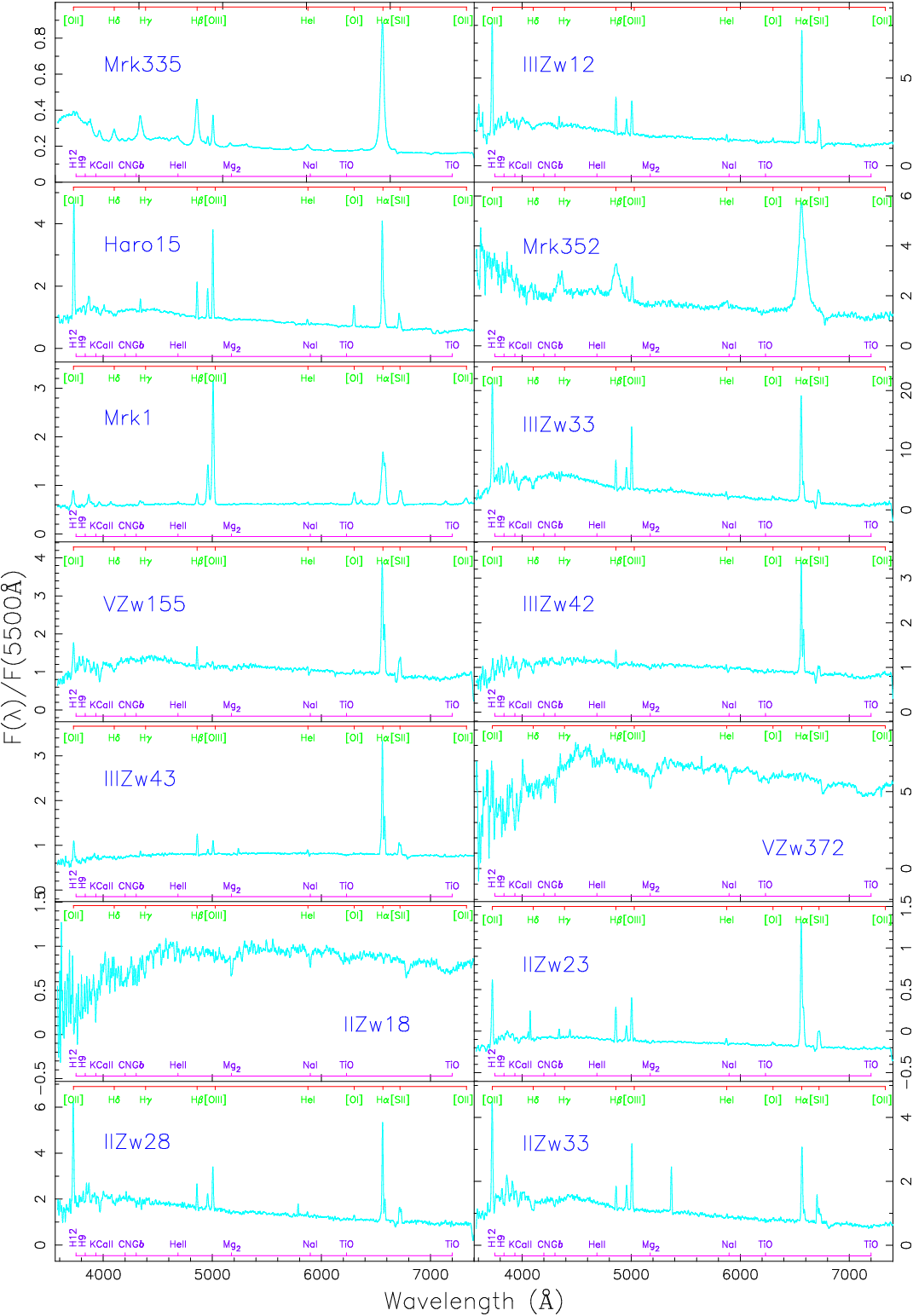}
\caption{
The atlas of rest-frame spectra in the electronic form.  The galaxies 
have been ordered according to their right ascension at epoch 2000.
Fluxes are normalized to a 50\AA\ region centered on 5500\AA. The top 
labels (for emission line galaxies) or the bottom labels (for absorption 
line galaxies) list the galaxy name. A horizontal bar is drawn at the 
top/bottom of the figure with the location of typical features in 
emission/absorption lines.
}
\label{fig4}
\end{figure*}
\setcounter{figure}{3}
\begin{figure*}
\includegraphics[width=\textwidth]{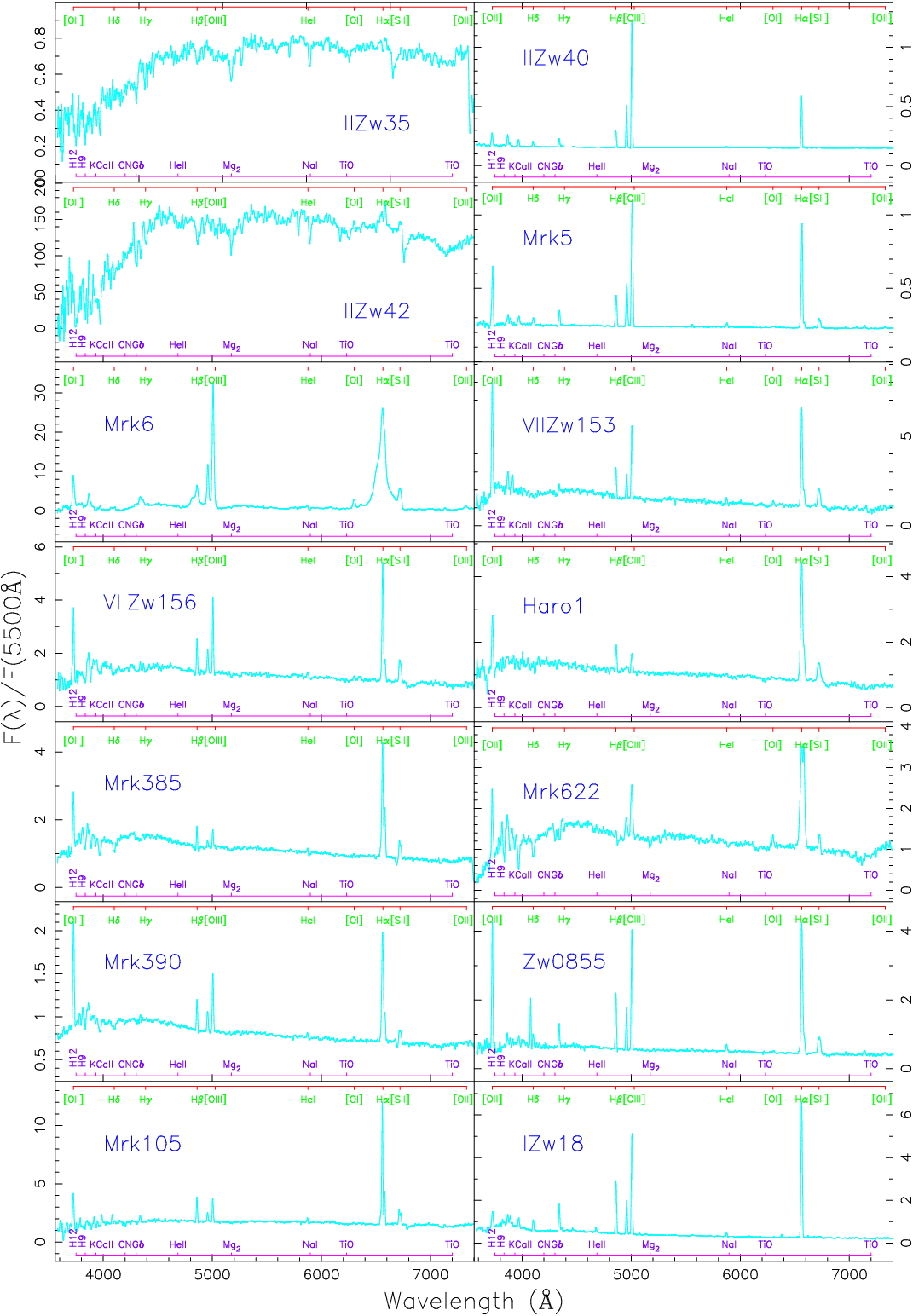}
\caption{\it Continued}
\end{figure*}

\setcounter{figure}{3}
\begin{figure*}
\includegraphics[width=\textwidth]{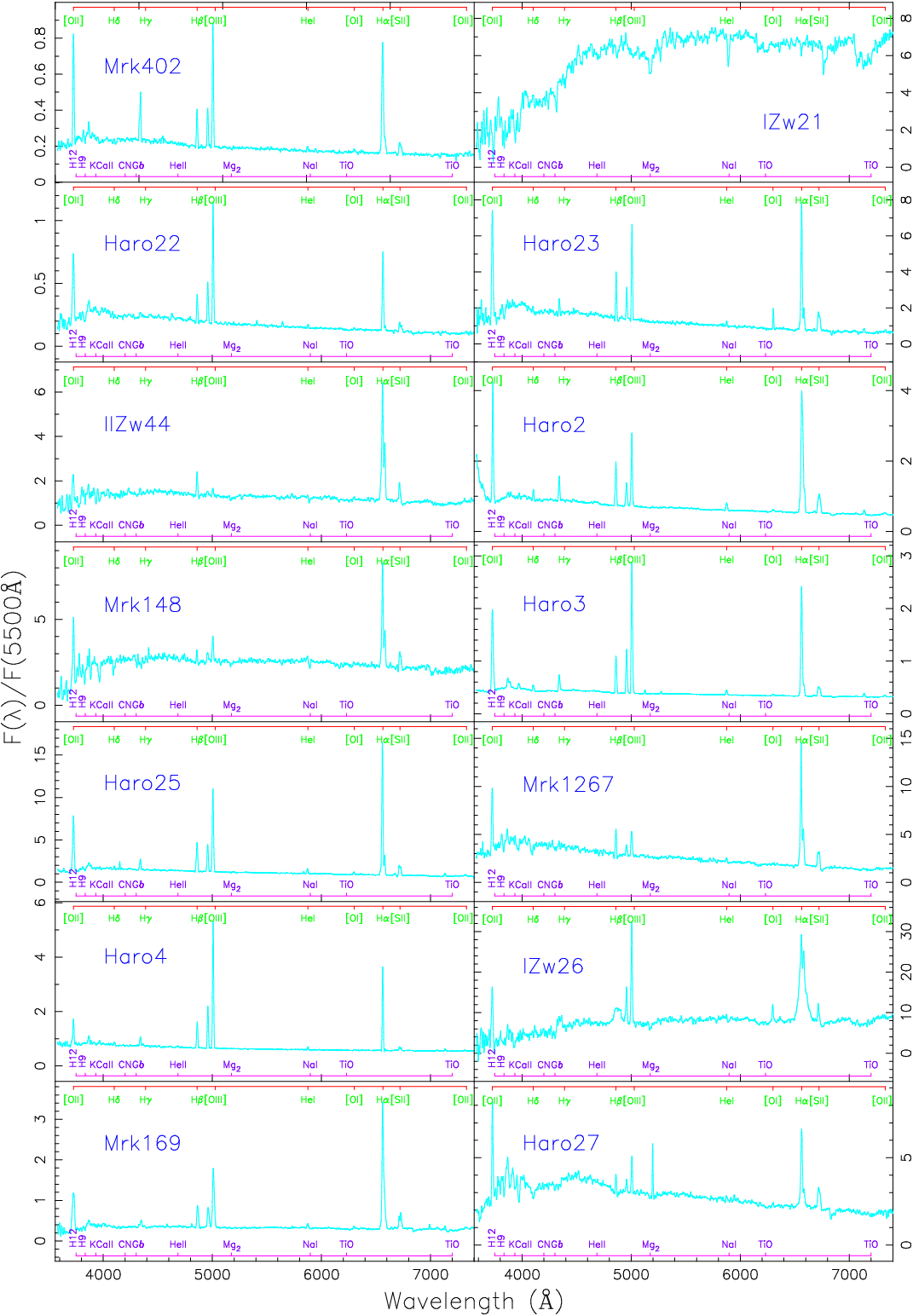}
\caption{\it Continued}
\end{figure*}
\setcounter{figure}{3}
\begin{figure*}
\includegraphics[width=\textwidth]{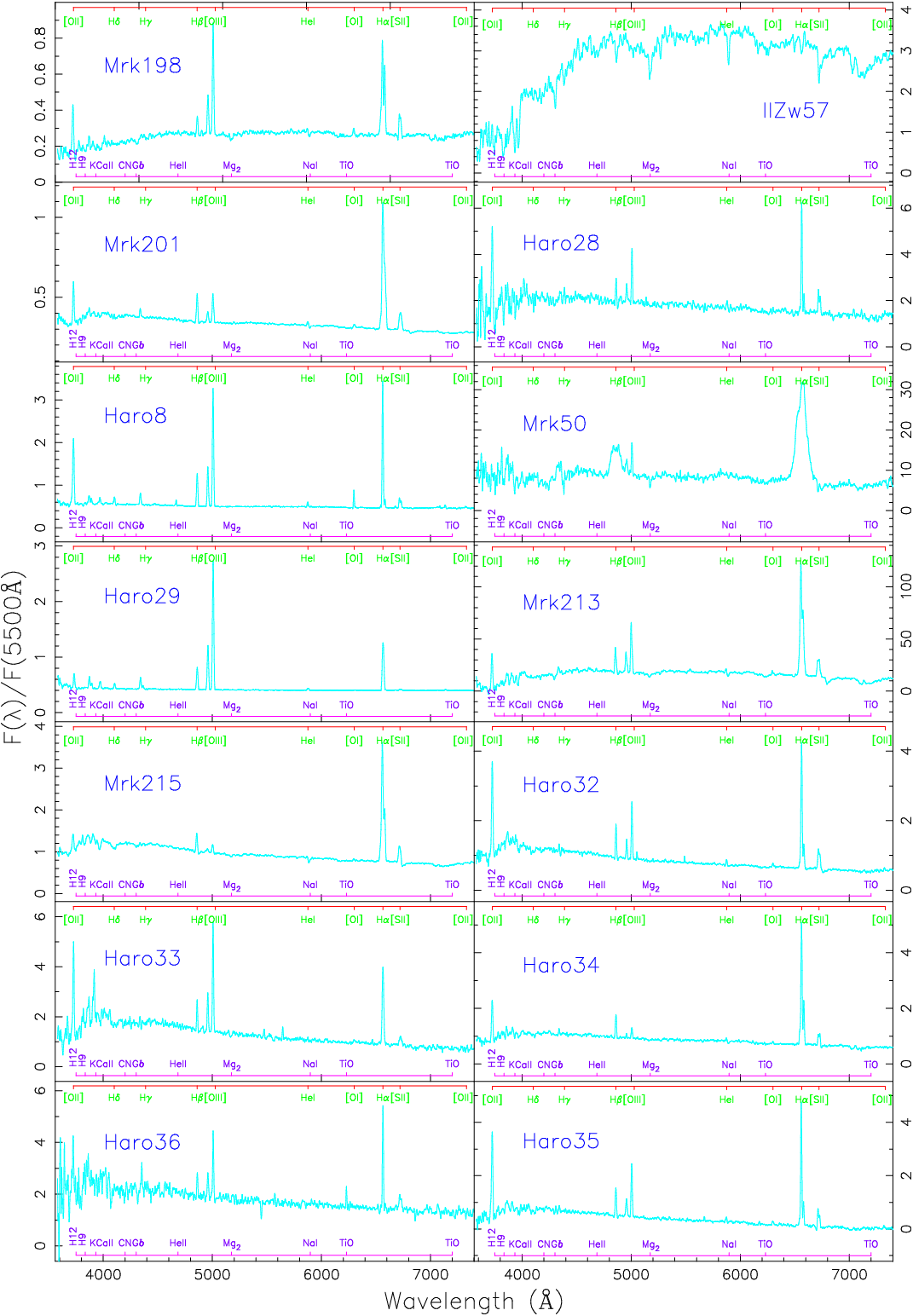}
\caption{\it Continued}
\end{figure*}
\setcounter{figure}{3}
\begin{figure*}
\includegraphics[width=\textwidth]{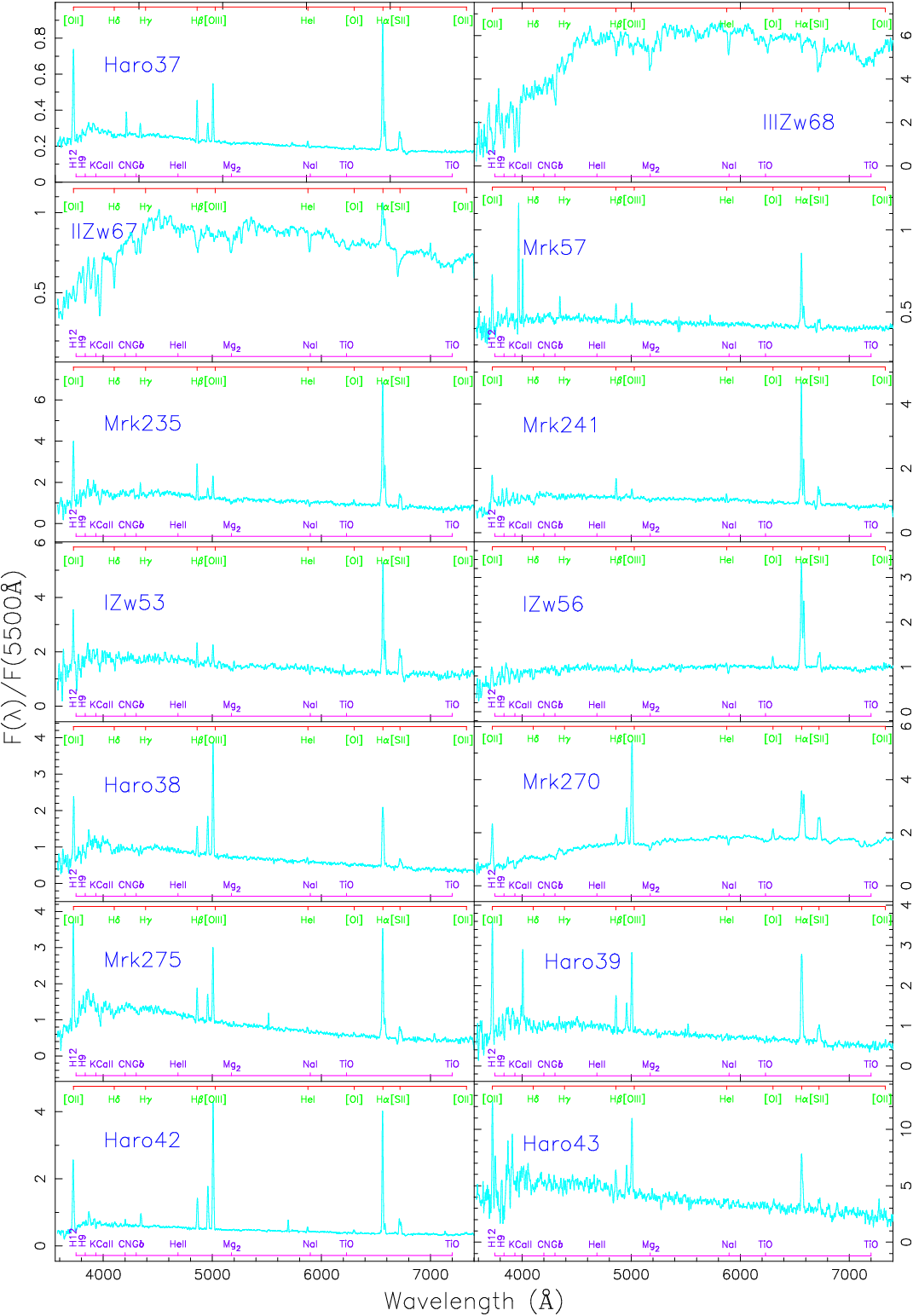}
\caption{\it Continued}
\end{figure*}
\setcounter{figure}{3}
\begin{figure*}
\includegraphics[width=\textwidth]{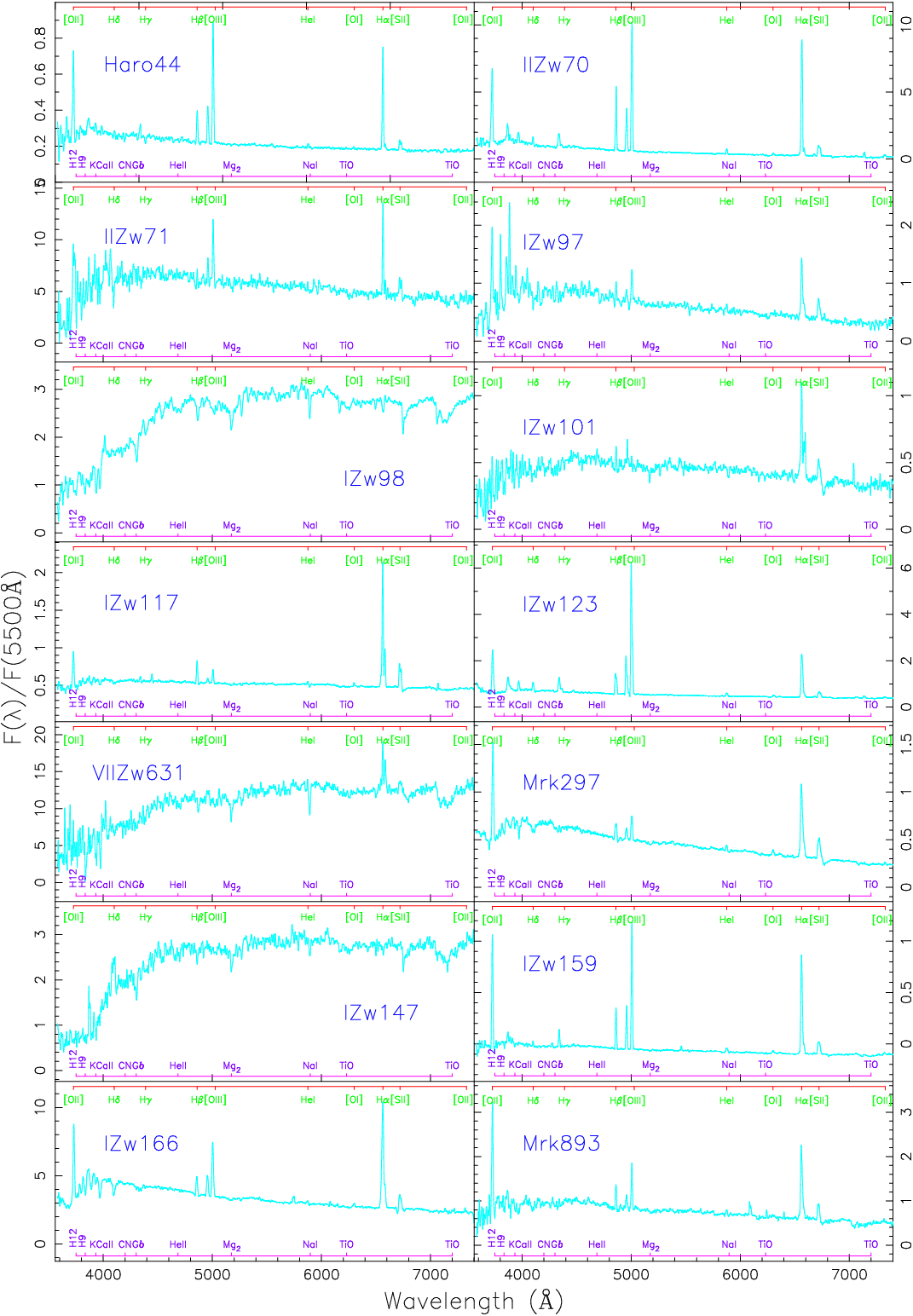}
\caption{\it Continued}
\end{figure*}
\setcounter{figure}{3}
\begin{figure*}
\includegraphics[width=\textwidth]{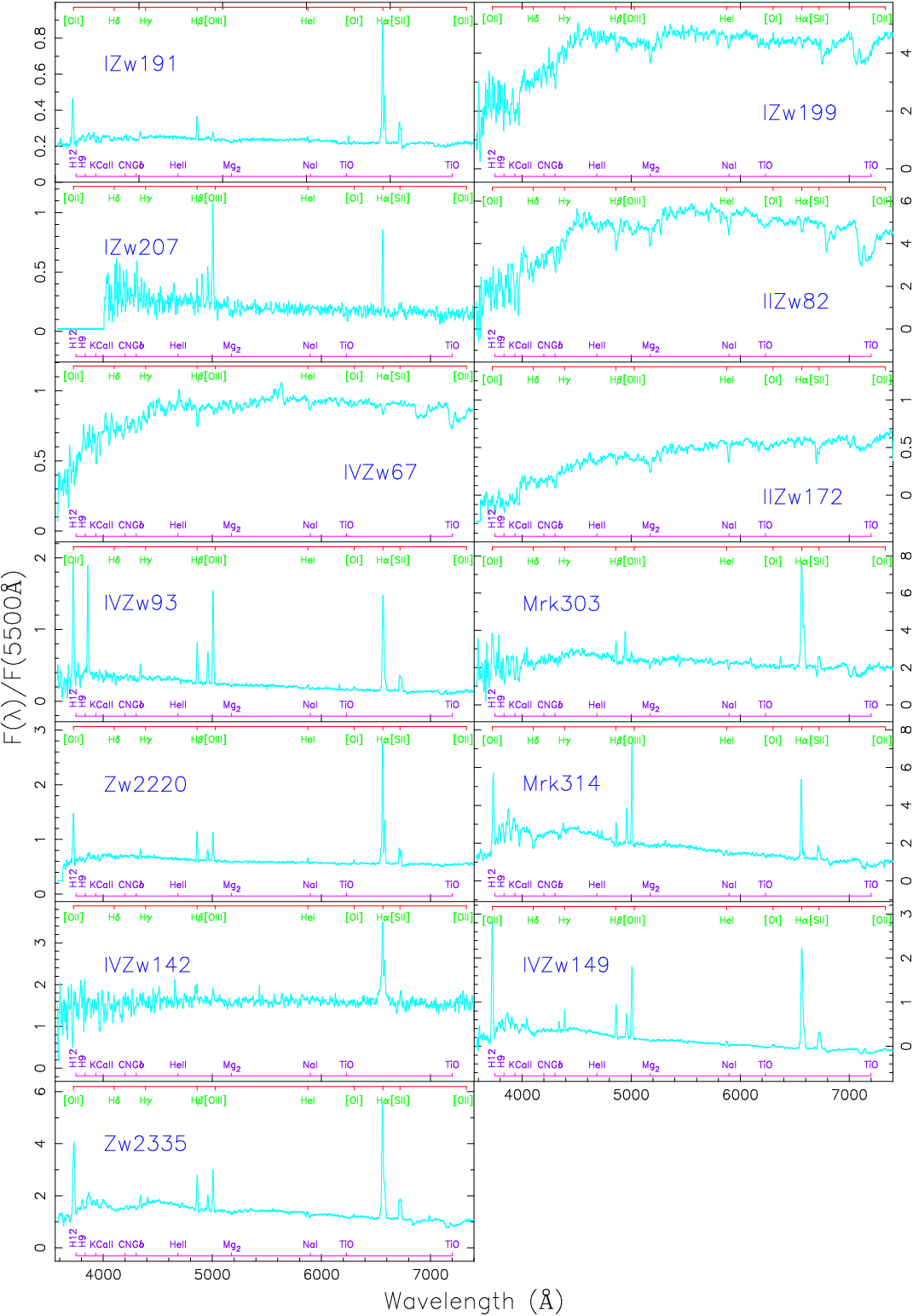}
\caption{\it Continued}
\end{figure*}

\section{Discussion}

\subsection{Spectrophotometric accuracy}

The spectrophotometric accuracy can be estimated by adding 
individual contributions to the total error in quadrature.  

The main sources of errors that affect the spectral shape include: 
the fitting of the sensitivity function, the published standard star 
fluxes, the adopted atmospheric extinction curve (\cite{jansen00}).
The error in the fit of the sensitivity function is 
estimated using the residuals of individual standard stars from the 
mean calibration.  These residuals are dominated by systematic 
differences in the sensitivity function fitted to the different 
stars, and are less than 5\%. Standard star fluxes are accurate to better 
than 3\% (\cite{massey88}).  Considering this 3\% uncertainty, 
we find that errors in our sensitivity function fits are likely to 
be less than 5\%.  Application of the BAO mean atmospheric 
extinction curve to correct our data introduces an error in the 
continuum slope of the spectra.  Because most galaxies were 
observed at low airmasses, we expect the error in the continuum slope 
to be less than 5\% at any wavelength.

The main sources of errors that affect restricted ranges in wavelength 
include: the flat field variations, wavelength calibration, the 
sky subtraction and residuals due to cosmic ray hits (\cite{jansen00}).
Differences between flat fields taken on different nights 
within a run are small. The contribution of the read noise to the 
error in the flat field is negligible.  Errors in the wavelength 
calibration introduce small errors in the inferred flux densities 
on scales comparable to the distance between individual calibration 
lines.  Errors in the dispersion solution are less than 0.3\AA.  
These dispersion errors produce spectrophotometric errors of at 
most 3\%.  Sky subtraction errors dominate the total error on small 
scales. Because BCGs are compact objects, errors from sky 
subtraction are less than 5\%.  Cosmic ray residuals introduce 
large errors in the extracted spectra only near emission lines, 
where the steepness of the local background renders a clean fit 
difficult. Residuals in continuum or sky portions in the spectra 
are smaller than 2\% of the local background. Errors may be as large 
as 10\% per extracted pixel. 

This analysis shows that the spectrophotometry is accurate to better 
than 10\% over small wavelength regions, and about 15\% or better 
on large scales.

\subsection{Internal checks of the spectrophotometry}

In Table 3 we list the total number of observations of each galaxy, 
39 galaxies have been observed twice. To check how consistent 
our calibration procedure is from night to night, including 
whatever errors exist in the adoption of a mean extinction curve, 
we can compare the final spectrophotometry to each individual 
observation. 
 
To illustrate this level of accuracy, we show in Figure 5a. duplicate 
spectra for three kinds of galaxies: an emission line galaxy (VIIZw 153), 
an absorption line galaxy (IIZw 35), and a Seyfert galaxy (Mrk335).  
The dates at which the spectra were obtained, the exposure times, and 
the effective airmasses during the observations are indicated. As expected, 
the deviations are rarely larger than 10\% from 4500 to 6800\AA.
We see that the average spectral energy distribution precision of our 
data is good. No systematic differences are seen.

\setcounter{figure}{4}
\begin{figure}
\includegraphics[width=\columnwidth]{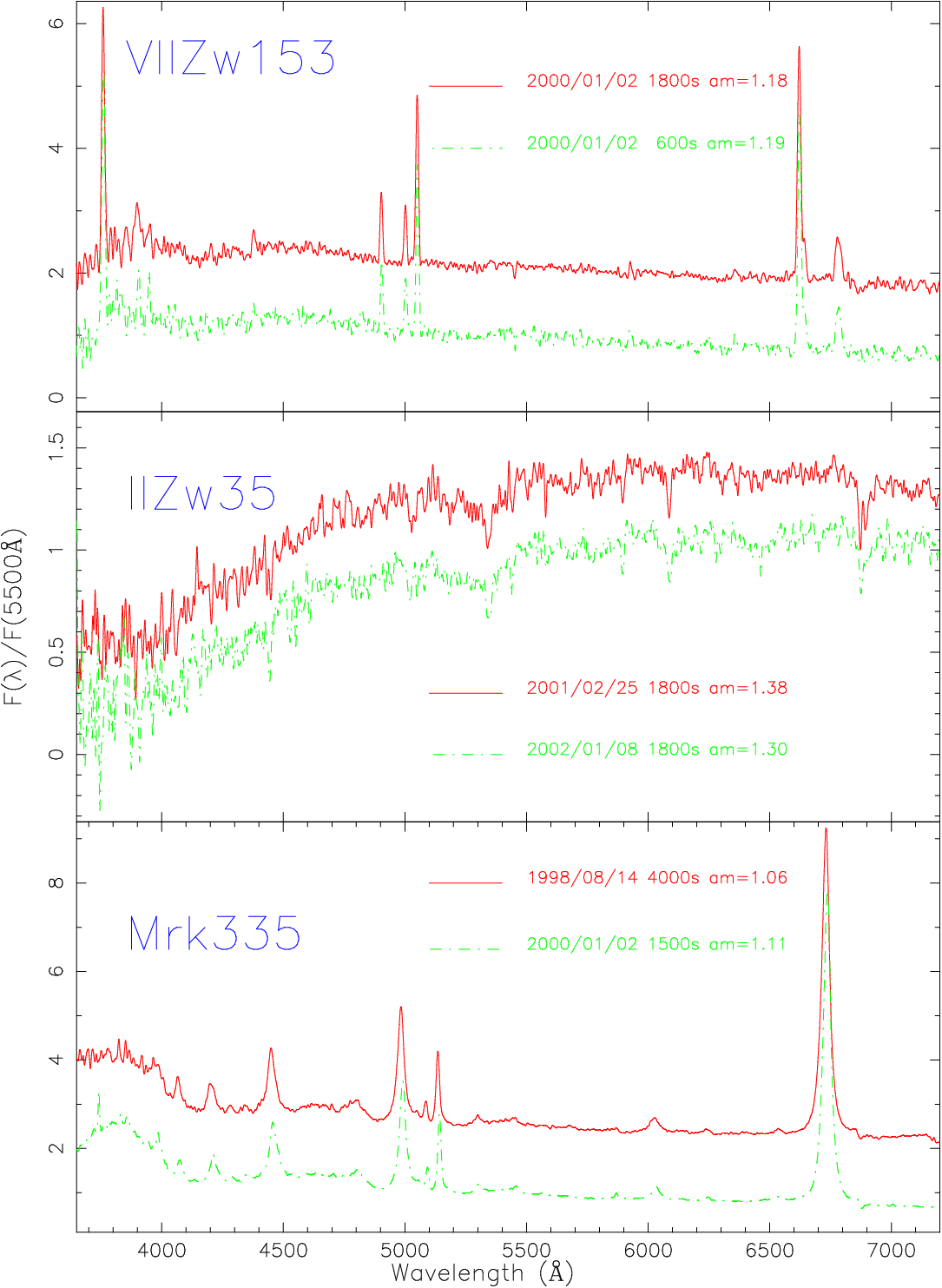}
\caption{{\bf a)}
A check of the internal consistency of our spectrophotometry using
the spectra of the 3 galaxies that we observed during two different
times. VII Zw 153: emission line galaxy; II Zw 35: absorption line
galaxy; Mrk 335: Seyfert galaxy.
All spectra were normalized to the average level in interval
5475 -- 5525\AA. The spectra in each panel has been offset by 0.2 for
clarity.}
\label{fig5a}
\end{figure}

\subsection{Comparison with other spectra}

Our sample of blue compact galaxies contains two galaxies that were 
also observed by Kennicutt (1992): Haro 3 and Mrk 201. We compare our
spectra for these objects with those obtained by Kennicutt (1992) in
Figure 5b.

In the 4500--6800\AA\ region the continuum of our spectra match 
Kennicutt's to better than 10\% over small ranges.  Bluewards of 
4500\AA\ differences tend to become larger, up to about 30\%.  In 
addition, our spectra show a bluer optical spectrum and stronger emission 
lines than do Kennicutt's. The aperture used by Kennicutt is much 
larger than ours (45$^{\prime\prime}$ circular versus 
3$^{\prime\prime}$ slit). Since BCGs have a central burst 
of star formation, the difference in aperture size is most likely 
the cause of the difference in continuum shape and emission lines. 
The increased aperture size allows for a greater contribution 
to the flux by older stars surrounding the central brust.

\setcounter{figure}{4}
\begin{figure}
\includegraphics[angle=-90,width=\columnwidth]{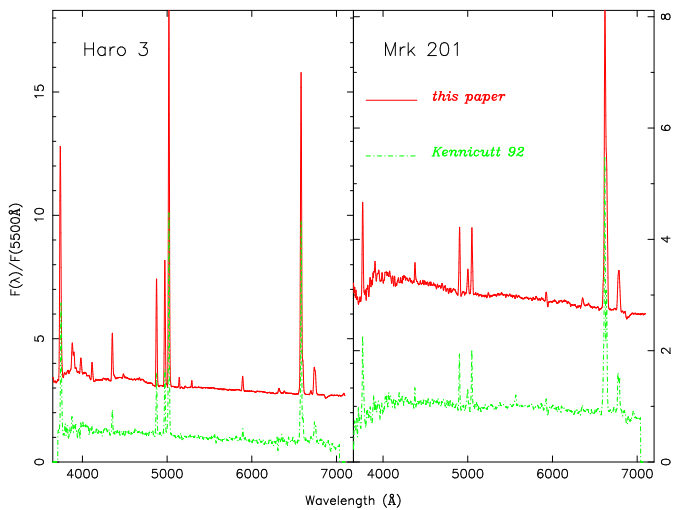}
\caption{{\bf b)} Comparison of our spectrophotometry with
Kennicutt (1992): Haro 3 and Mrk 201. The spectra have all been
normalized to the flux centered on 5500\AA\ and have been offset
by 0.2 for clarity.}
\label{fig5b}
\end{figure}

\section{Summary}

We have presented medium-resolution spectroscopy observations 
of 97 blue compact galaxies.  These BCGs cover a range in absolute 
blue magnitude $-21.9 < M_B < -14.0$; 83 BCGs have $M_B < -17$ mag.  
The spectra were obtained with the Beijing Astronomical Observatory 
2.16 m telescope, with spectral resolution (FWHM) of about 10{\AA} 
and spectral range 3580{\AA} -- 7400{\AA}. The spectrophotometry is 
expected to be accurate to 10\%.

Based on the emission and absorption lines, we measured the recession 
velocities of the sample galaxies. Most galaxies in the sample have 
velocities below 9000 km s$^{-1}$; the error in recession velocity is 
less than 67 km s$^{-1}$.

The majority of the sample galaxies are dominated by active star 
formation; 74 of them have typical \hii\ regions spectra, the 
others are Seyfert galaxies and absorption line galaxies.

Measurement of the spectral line strengths and the results of the 
spectral analysis of these galaxies will be presented in the next 
paper of this series. We will use these data to understand the star 
formation history, the physical parameters governing the burst mode 
of star formation and chemical evolution in blue compact galaxies.
 
\begin{acknowledgements}

We are grateful to Dr. A. Weiss and Dr. S. Charlot for their 
helpful comments, constructive suggestions, and hard work of
English revision for this paper. The referee Dr. D. Kunth is
thanked for many useful suggestions.
We also thank the BAO Chinese 2.16m Telescope time allocation 
committee for their support of this programme and to the staff 
and telescope operators of the Xinglong Station of the Beijing 
Astronomical Observatory for their support. Especially we would 
like to thank Dr. J.Y. Wei and Dr. X.J. Jiang for their active 
cooperation that enabled all of the observations to go through 
smoothly.  
This work is supported by the Chinese National Natural Science 
Foundation (CNNSF 10073009).  
Dr. X. Kong has been financed by the Special Funds for Major State 
Basic Research Projects of China and the Alexander von 
Humboldt Foundation of Germany.
\end{acknowledgements}

\end{document}